\documentclass[twocolumn,showpacs,preprintnumbers,amsmath,amssymb]{revtex4}

\usepackage{color}

\usepackage{graphics}
\usepackage{graphicx}
\usepackage{dcolumn}
\usepackage{bm}
\usepackage{amssymb}
\pdfoutput=1

\begin{document}
\preprint{APS/123-QED}

\title{Charge, spin and lattice effects in the spin-Peierls ground state of MEM(TCNQ)$_2$}
\author{Mario Poirier }
\author{Mathieu de Lafontaine }
\author{Claude Bourbonnais }
\affiliation{  Regroupement Qu\'eb\'ecois sur les Mat\'eriaux de Pointe, D\'epartement de Physique, Universit\'e de Sherbrooke,
Sherbrooke, Qu\'ebec,Canada J1K 2R1}
\author{Jean-Paul Pouget}
\affiliation{ Laboratoire de Physique des Solides, CNRS UMR 8502, Universit\'e Paris-Sud, 91405 Orsay C\'edex, France}

\date{\today}

\begin{abstract}
We report an investigation of charge, spin and lattice effects in the spin-Peierls state of the organic compound MEM(TCNQ)$_2$. The 16.5 GHz dielectric function along the chain axis shows an enhancement below the spin-Peierls transition temperature near 18 K  consistent with the charge coupling to the elastic strain involved in the transition. The velocity of two elastic modes perpendicular to the chain axis presents anomalies at the transition which can be explained with a Landau free energy model including a linear-quadratic coupling energy term between the appropriate elastic strain $e$ and the spin-Peierls magnetic gap $\Delta_q$. The analysis of the dielectric and elastic features aims toward an order parameter with an associated critical exponent $\beta \sim$ 0.36, which is  similar to the  three-dimensional behavior  seen in other spin-Peierls materials. All these effects studied in a magnetic field up to 18 Teslas appear also compatible  with a mean-field model of a quasi-one-dimensional spin-Peierls system.

\end{abstract}

\maketitle
\section{INTRODUCTION}
In antiferromagnetic  spin chains, the spin-Peierls (SP) transition results from a coupling between the electronic spin correlations  and phonons yielding a dimerized ordered singlet state below a transition temperature $T_{\rm SP}$. The SP state has  been formerly investigated in a few organic materials  (TTF-CuBDT, TTF-AuBDT, MEM(TCNQ)$_2$\cite{Bray1973,Bray1983},  and more recently in an inorganic compound CuGeO$_3$\cite{Hase1993}. Despite different crystals structures, very similar physical properties are observed. For example, the critical exponent $\beta$ associated to the onset of the SP order parameter was found to be consistent with a 3D Ising universality class in both CuGeO$_3$ \cite{Lumsden1998} and MEM(TCNQ)$_2$ \cite{Lumsden1999}. The magnetic phase diagram also shows a common dimerized (D)-uniform (U) boundary for all these compounds \cite{Hase1993}, in agreement with the theoretical prediction \cite{Cross1979}.

Alongside, the SP state has also been found in   series of quasi-1D  ion-radical organics salts, including the (BCPTTF)$_2 X$ \cite{Liu1993,Dumoulin1996},  and the  Fabre series (TMTTF)$_2X$  \cite{Pouget82,Creuzet1987},  both with $X$ =PF$_6$ and AsF$_6$. In the (TMTTF)$_2$PF$_6$ compound for instance, the transition takes place at $T_{\rm SP} \approx 18$ K at ambient pressure for  the salt with hydrogenated TMTTF molecule \cite{Creuzet1987,Foury04,Chow98} and at 13 K in the deuterated one \cite{Pouget2006}. These quasi-1D materials, however, display a slightly dimerized structure leading to a  Mott insulating gap below $T_{\rho} \approx 230$K \cite{Coulon82}, followed by a charge ordered superstructure (CO) at $T_{CO}\approx 65 $K \cite{Chow00,Nad2000}($T_{CO}\approx 84$K in deuterated \cite{Coulon07,Langlois2010})  whose proximity with the SP transition  has prevented  the full characterization of its order parameter. Indeed, inhomogeneous dielectric \cite{Langlois2010} and elastic \cite{Poirier2012} behaviors were observed in the vicinity of the SP transition for the hydrogenated and deuterated PF$_6$ salts. Although the magnetic field dependence of the SP transition temperature agrees satisfactorily with the predicted one (D-U boundary), the field dependence of the order parameter appears to differ from the theoretical prediction \cite{Azzouz1996}, at variance with   the inorganic CuGeO$_3$ material \cite{Poirier1995}.

The organic compound MEM(TCNQ)$_2$ is composed of planar TCNQ molecules which stack along the crystallographic $\textit{\textbf{c}}$ axis producing a quasi-one-dimensional chains structure \cite{Bodegom1981}. Below 335 K, the compound undergoes a metal-insulator transition that dimerized the TCNQ chains \cite{Huizinga1979}, but leaves the spins degrees of freedom unaffected down to $T_{SP}\approx 18$K, where a second transition tetramerizes the chains and shows all the characteristics of the SP distorted state \cite{Bodegom1981a}. However, not much is known on the SP gap in this system \cite{Blundell1997,Lumsden1999}. Moreover, deviations from the universal structure of the magnetic phase diagram have been  reported \cite{Weckhuysen2001}. In contrast   to the (TMTTF)$_2$PF$_6$ compound, which has also a triclinic structure, there is no CO ordering or any known transition between 18 and 335 K in MEM(TCNQ)$_2$, allowing in principle a better characterization of its SP ordered state.

In this paper we investigate charge, spin and lattice effects on the SP transition in MEM(TCNQ)$_2$. Charge effects are analyzed by measuring the microwave dielectric constant at 16.5 GHz, whereas spin and lattice effects are studied by measuring the magneto-elastic coupling at ultrasonic frequencies. An increase of the dielectric constant is observed below $T_{\rm SP}$, consistently with the building-up of the SP order parameter. Anomalies on two elastic moduli are observed at the SP transition. All these effects are studied as a function of temperature and magnetic field.

\section{EXPERIMENT}
The MEM(TCNQ)$_2$ compound has a triclinic crystal structure and single crystals grow as long slabs from which small pieces having typical dimensions 2.4x0.8x0.4 mm$^3$ are cut. The long axis corresponds to the stacking direction $\textit{\textbf{c}}$ and the two large parallel faces correspond to the $\textit{\textbf{a$^\prime$ c}}$ plane ($\textit{\textbf{a$^\prime$}}$ being the component of $\textit{\textbf{a}}$ perpendicular to $\textit{\textbf{c}}$) whose normal direction is ${\textit{\textbf{b}}}^*$. We used a standard microwave cavity perturbation technique \cite{Buravov} to measure the complex dielectric function $\epsilon^* = {\epsilon}_{1} + i{\epsilon}_{2}$ along the $\textit{\textbf{c}}$ axis.  A copper cavity resonating in the TE$_{102}$ mode was used at 16.5 GHz. The organic slab is inserted in a quartz capillary tube and immobilized by thin cotton threads to prevent any stress or movement during thermal cycling. The tube is then glued on a quartz rod to allow its insertion in the cavity and the precise orientation of the slab along the microwave electric field. Following the insertion of the sample, changes in the relative complex resonance frequency $\Delta f/f + i \Delta(1/2Q)$ ($Q$ is the cavity quality factor) as a function of temperature are treated according to the depolarization regime analysis after substraction of the tube contribution.

We use a pulsed ultrasonic interferometer to measure the variation of acoustic velocity along ${\textit{\textbf{b}}}^*$ relative to its value at a fixed temperature $T_0$ = 25 K, $\Delta V / V$ = $[V(T) - V(T_0)] / V(T_0)$; only the direction ${\textit{\textbf{b}}}^*$ corresponding to natural parallel faces of the crystal can be investigated. The ultrasonic technique is used in the transmission mode and, because the typical thickness of the crystal along the ${\textit{\textbf{b}}}^*$-axis is quite small around 0.4 mm, a CaF$_2$ delay line must be used to separate the first transmitted acoustic echo from the electric pulse. The acoustic pulses are generated with LiNbO$_3$ piezoelectric transducers resonating at 30 MHz and odd overtones bonded to the crystals with silicone seal. Since the crystal structure is triclinic, one acoustic mode is characterized as quasi-longitudinal and another as quasi-transverse (polarization along ${\textit{\textbf{a$^\prime$}}}$). Because the velocity is related to the mass density $\rho$ and a particular elastic modulus $C_{ij}$ modulus through the relation $C_{ij} = \rho V^2$, the $\Delta V / V$ data are directly the image of the relative variation of the compressibility modulus $C_{22}$ (longitudinal) or the shear modulus $C_{66}$ (transverse), if we ignore the density variations in first approximation. Because of unavailability, the microwave and the ultrasonic experiments were performed on only one crystal which had appropriate dimensions. A magnetic field up to 18 Tesla could be applied along the ${\textit{\textbf{a$^\prime$}}}$-axis. Unless otherwise stated in the discussion, all velocity data were obtained for a frequency around 100 MHz.

\begin{figure}[H,h]
\includegraphics[width=8.5cm]{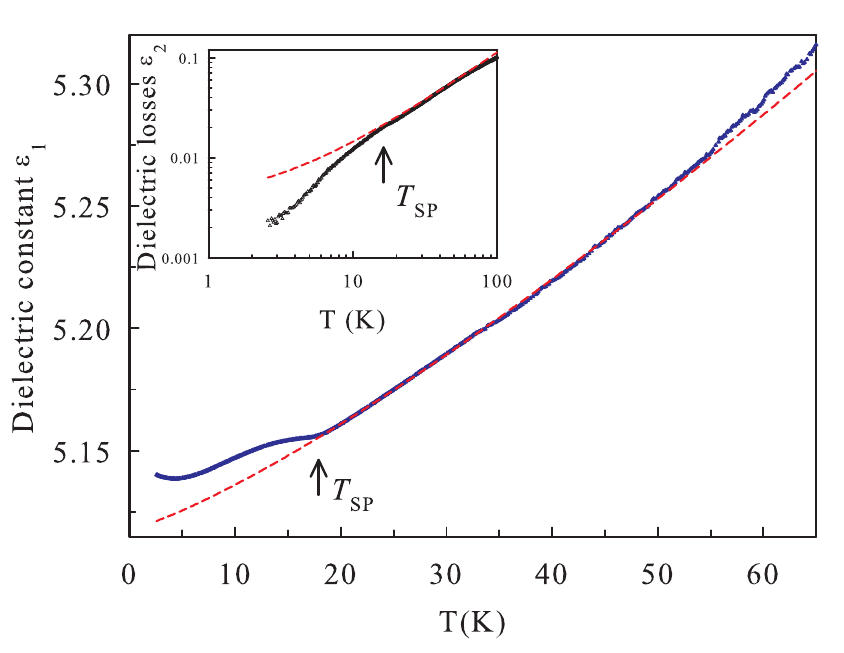} \caption{(Color online)
Temperature dependence of the dielectric constant along the chain axis below 60 K. Inset: dielectric losses  on a log-log scale. The dashed curves represent a polynomial extrapolation of the background below 50 K.}\label{fig.1}
\end{figure}

\section{RESULTS AND DISCUSSION}
\subsection{Dielectric constant}
In the insulating state below 335 K, the dielectric constant at microwave frequencies (9 GHz) was reported in a previous work to be almost constant with an absolute value between 10 and 16 \cite{Morrow1980}. Although a small decrease was observed below 60 K, no anomalous behavior at the SP transition was reported. We present in Figure~1 the temperature dependence of the dielectric constant $\epsilon_1$ below 60 K. The absolute value is just above 5 and it decreases weakly with temperature in a very smooth manner down to 18 K where a small enhancement is observed; the dielectric losses $\epsilon_2$ show also a change in the decreasing rate below the same temperature (inset Fig.1). The $\epsilon_1$ enhancement coincides with the onset of SP order below 18 K. To analyze more precisely this enhancement, the insulating state between 20 and 50 K has been tentatively extrapolated to 2 K using a second degree polynomial fit, as shown by the dashed curve. The substraction between the $\epsilon_1$ data and this extrapolation curve below 20 K gives the  $\Delta\epsilon$ enhancement shown in Figure~2 for different magnetic field values.

\begin{figure}[H,h]
\includegraphics[width=8.5cm]{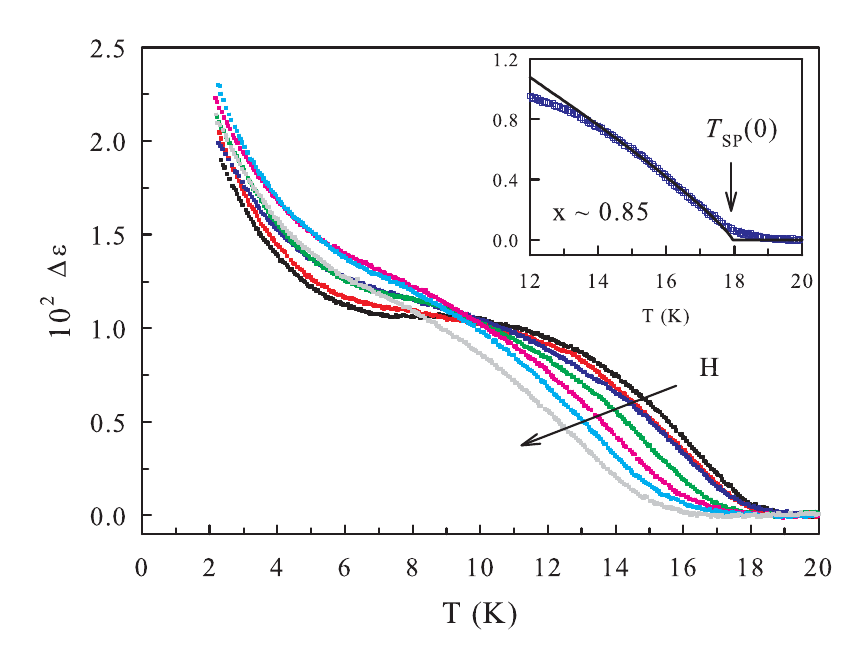} \caption{(Color online)
Dielectric constant enhancement $\Delta\epsilon$ as a function of temperature below 20 K for different magnetic field values: 0, 4, 8, 12, 16, 17 and 18 Teslas. Inset: critical behavior in zero field fitted to the power law $(1- T/T_{\rm SP})^x$ (continuous line).}\label{fig.2}
\end{figure}

In zero magnetic field, $\Delta\epsilon$ increases rapidly with decreasing temperature below 18 K, saturates near 9 K before increasing further down to 2 K. The fast increase below 18 K and the saturation near 9 K appear to mimic the growth of the SP order parameter as previously observed in hydrogenated (TMTTF)$_2$PF$_6$ \cite{Langlois2010}; for MEM(TCNQ)$_2$, however, the growth is much slower than for (TMTTF)$_2$PF$_6$. If we fit $\Delta\epsilon$ enhancement to a power law $(1- T/T_{\rm SP})^x$ near the critical temperature $T_{\rm SP}$ = 17.9 K (inset, Fig.2), we obtain $x \simeq 0.85(5)$. If we assume a linear relationship between $\Delta \epsilon$ and the SP order parameter, $x$ is associated to the related critical exponent $\beta$; its value  is significantly larger than the  one  expected for a SP - one component - order parameter in three dimensions (3D), a value observed in similar dielectric experiments for (TMTTF)$_2$PF$_6$ ($\beta \simeq 0.36$) \cite{Langlois2010}. This is surprising since MEM(TCNQ)$_2$ is thought to be more isotropic, as shown by the lattice fluctuations precursor to the transition, which are three dimensional in character \cite{Liu1993,Bodegom1981a}. Moreover, an exponent consistent with a 3D behavior was reported for the X-ray scattering intensity $I\sim (1- T/T_{\rm SP})^{2\beta}$ for MEM(TCNQ)$_2$ with $\beta \simeq 0.35(6)$ \cite{Lumsden1999}. If one rather assumes that the  dielectric response $\Delta \epsilon$ also scales with the square of the SP order parameter this would yield $x = 2\beta$, a value congruent with the one reported in X-ray experiments \cite{Lumsden1999}.  At saturation near 9 K, the relative amplitude of $\Delta\epsilon$ (0.2\%) is much smaller compared to (TMTTF)$_2$PF$_6$ (0.9\%). This is likely due to the larger polarizability of the latter material caused by the ferroelectricity associated to the CO ordering above $T_{\rm SP}$.

The further increase of the dielectric constant below 9 K does not appear to be related to the SP ground state. It rather shows a Curie type temperature dependence  possibly due to local electric dipoles. The total  dielectric constant enhancement at low temperatures is the sum of two terms, the SP contribution $(\Delta\epsilon)_{\rm SP}$ just discussed and a Curie tail of the form, $A/T$,
\begin{equation}
\Delta\epsilon = (\Delta\epsilon)_{\rm SP} + A/T
\end{equation}
This assumption is further supported by the magnetic field effects reported in Figure~2. As expected, a magnetic field (oriented along the ${\textit{\textbf{a$^\prime$}}}$ axis) shifts downward  the SP temperature $T_{\rm SP}$ and consequently modifies the overall temperature dependence of $\Delta\epsilon$. Saturation near 9 K is progressively replaced by a variation of the increasing rate. The temperature dependence of the Curie term does not appear to be modified by the magnetic field although the absolute value of $\Delta\epsilon$ is increasing with field (the 18 Tesla curve excepted). This increase is likely due to a magnetic field dependent $(\Delta\epsilon)_{\rm SP}$ at low temperatures as previously observed for (TMTTF)$_2$PF$_6$ \cite{Langlois2010}, and not a field dependent constant $A$ of the Curie term. A modification of the coupling between the charge and the SP lattice distortion in a presence of a magnetic field, and not the zero-temperature value of the SP order parameter, may be responsible for this slight increase of polarizability at low temperatures. Although the origin of the low temperature Curie tail in the dielectric response remains to be clarified, it should be mentioned that in the same temperature range a Curie tail  has also been reported in the spin sector for the magnetic susceptibility \cite{Huizinga1979}.

For these dielectric experiments, the determination of the zero field transition temperature $T_{\rm SP}(0)$ is particularly difficult because of the slow power law in the reduced temperature ($x \simeq 0.85$) and the presence of substantial critical scattering below and above the transition. Following the example of X-ray diffraction experiments \cite{Lumsden1999}, the second  derivative of $(\Delta\epsilon)_{\rm SP}$ with respect to $T$ yields the best determination of $T_{\rm SP}(0)$, which is also consistent with the ultrasonic velocity data shown in the next section. From Figure~2, we notice that a magnetic field increases the width of the critical domain without changing much the exponent $x$.

\subsection{Ultrasonic velocity}
\begin{figure}[H,h]
\includegraphics[width=8.5cm]{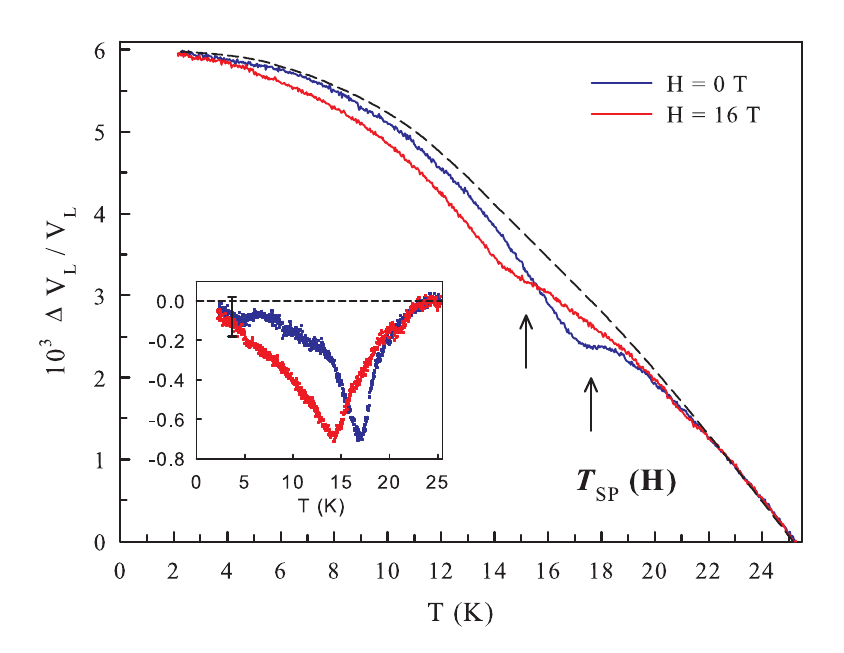} \caption{(Color online)
Temperature dependence of the relative variation of the longitudinal velocity $\Delta V_L / V_L$ below 25 K at two magnetic field values, H = 0 and 16 Teslas. Arrows indicate the transition temperature $T_{\rm SP}$. The dashed curve mimics the elastic background. Inset: softening anomalies due to the SP transition after substraction of the background.}\label{fig.3}
\end{figure}

The velocity of the longitudinal and transverse  elastic modes investigated here increases monotonically with decreasing temperature from 200 K down to 20 K, without indication of an unusual elastic behavior. Below 20 K however, the SP transition produces a well defined anomaly that appears similar for both modes, although the amplitude is larger for the transverse one. We present in Figure~3 the temperature dependence of the relative variation of the longitudinal velocity $\Delta V_L / V_L$ below 25 K at the two magnetic field values, $H = 0$ and 16 Teslas. As the temperature is decreased below 25 K in zero field, $\Delta V_L / V_L$ increases until it shows a local maximum near 18 K, followed by a dip and then an increase that continues down to 2 K where it saturates. In constrast with the situation found in other inorganic \cite{Poirier1995} and organic \cite{Poirier2012} materials for an equivalent elastic modulus, the dip indicates that this particular mode, $C_{22}$, softens importantly near $T_{\rm SP}$ before its stiffening at lower temperature in the SP ordered state. A magnetic field of 16 Teslas widens and shifts the anomaly down to 15 K without affecting the $\Delta V_L / V_L$ data above $T_{\rm SP}$ and only very weakly (within the uncertainties) in the low temperature limit where saturation is observed. These results suggest that the normal elastic background for this particular mode is given by the dashed curve of Fig.~\ref{fig.3}. When this background is substracted from the data, we obtain the sound velocity variation shown in the inset of Fig.~\ref{fig.3}. If we locate the SP transition temperature near the  minimum, after some fluctuation precursors, the sound velocity displays a sudden softening at $T_{\rm SP}$, followed by a progressive stiffening in the SP ordered state that evolves  toward its normal phase value in the low temperature limit.

\begin{figure}[H,h]
\includegraphics[width=8.5cm]{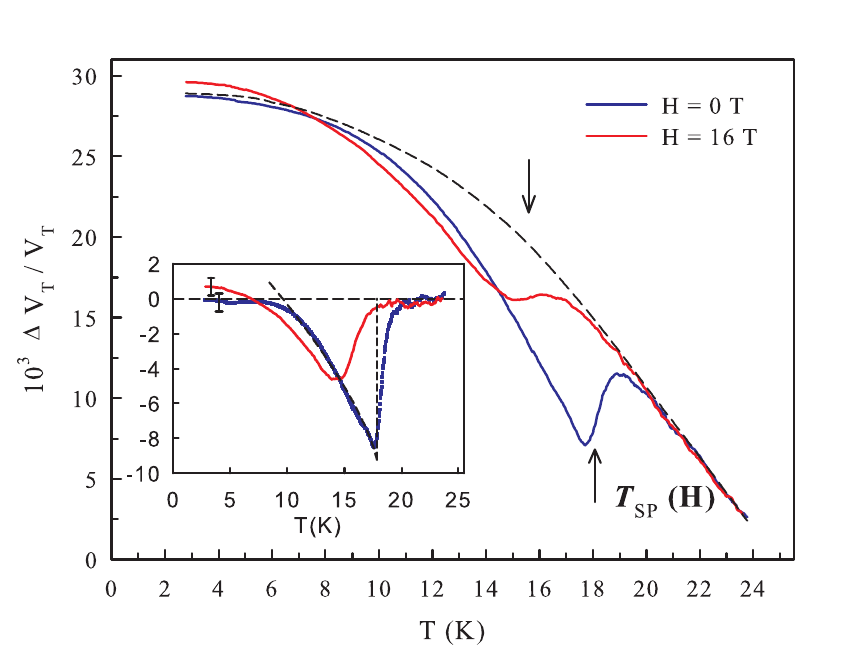} \caption{(Color online)
Temperature dependence of the relative variation of the transverse velocity $\Delta V_T / V_T$ below 25 K at two magnetic field values, H = 0 and 16 Teslas. Arrows indicate the transition temperature $T_{\rm SP}$. The dashed curve mimics the elastic background. Inset: softening anomalies due to the SP transition after substraction of the background; the dashed line is obtained with a $\Delta_q \sim (1- T/T_{\rm SP})^{0.36}$ critical behavior.}\label{fig.4}
\end{figure}

The velocity data for the transverse mode, $\Delta V_T / V_T$, are presented in Figure 4 over the same temperature range and for the same field values. As a whole, the temperature dependence is very similar to the previous one although the elastic modulus concerned here is mainly $C_{66}$, a modulus that has not been investigated for other SP systems. For this particular elastic mode, the anomaly is much sharper in amplitude at the SP transition. The effects of the magnetic field are also similar to the previous ones, namely a down shift in temperature and a broadening of the transition, along with the absence of field effects above $T_{\rm SP}$ and at the lowest temperatures below 4 K (within the uncertainties). Similarly to the longitudinal mode, we simulate the most appropriate elastic background of the normal phase from the dashed curve of the Fig.~\ref{fig.4}. The inset of Fig.~\ref{fig.4} shows the result of the substraction from the background. At the approach  of $T_{\rm SP}$, the larger amplitude and the sharper edge in comparison with the longitudinal case appears coherent with the implication of a transverse displacement of the TCNQ in their molecular plane in the SP lattice distortion, as revealed by neutron diffraction structural refinements \cite{Visser1983}. Here again, the anomaly in the sound velocity superimposes two effects, a sudden softening at $T_{\rm SP}$ followed by a progressive stiffening in the ordered phase. For both elastic modes, $T_{\rm SP}$ is determined from the maximum value of the first temperature derivative of the velocity data.

In other SP systems \cite{Poirier1995,Poirier2012}, softening and/or stiffening anomalies observed on elastic moduli below $T_{\rm SP}$ were analyzed in a phenomenological framework of the Landau free energy model. This is achieved by introducing a coupling term between the SP distortion of wave vector $\textbf{q}$, which is proportional to the magnetic gap $\Delta_q$, and the uniform elastic deformation $e$ \cite{Pouget1989},
\begin{equation}
\label{Fc}
F_c = h |\Delta_q|^2 e^r,
\end{equation}
$h$ being the coupling constant.  This term is added to the SP Landau free energy
\begin{equation}
\label{ }
F_{SP} = a(T)|\Delta_q|^2 + b|\Delta_q|^4 + f|\Delta_q|^6,
\end{equation}
where $a(T)=a'(T-T_{SP})$, and $ a', b,f>0$. As to the elastic energy, it is given by
\begin{equation}
\label{ }
F_{el} = C_{ii}^0 e^2/2,
\end{equation}
where $C_{ii}^0$ is the bare modulus. The minimization of the $F_{SP}+ F_c$ with respect to the gap $\Delta_q$ allows to calculate the equilibrium expression for the free energy, from which the renormalization of the elastic modulus  $ C_{ii} =\partial^2F_{tot}/\partial e^2_i $ can be extracted.

One can distinguish two cases: first, for a biquadratic coupling term  where $r$ = 2 in (\ref{Fc}), only a stiffening of the modulus is obtained below $T_{\rm SP}$, which is linear in $h$,
\begin{equation}
C_{ii} = C_{ii}^0 + 2 h |\Delta_q|^2.
\end{equation}
Although this type of stiffening anomaly on equivalent elastic moduli was observed for inorganic \cite{Poirier1995} and organic \cite{Poirier2012} SP compounds, it cannot explain the important softening seen at $T_{\rm SP}$ from  Figs. 3-4. In the second situation when a linear-quadratic ($r$ =1) coupling term is considered, a softening of the modulus proportional to the square of the coupling constant, -${{h^2}/{2b}}$, is  found. Moreover, when     the sixth order term $f|\Delta_q|^6$ is taken into account in  the free energy, it is superimposed to a stiffening contribution due to the onset of a finite $|\Delta_q|$ below $T_{\rm SP}$ simulating the decrease  of spin degrees of freedom responsible for the screening of the elastic modulus. This contribution  is also proportional to $h^2$, which gives
\begin{equation}
C_{ii} = C_{ii}^0 - {{h^2}\over{2b}} + {{3h^2f|\Delta_q|^2}\over{2b^2}}.
\label{Cii}
\end{equation}
As we have seen  in Figs. 3-4 for MEM(TCNQ)$_2$, the stiffening is sizeable since the normal elastic behavior is, within the  error bar, practically restored at low temperatures. For comparison, in the CuGeO$_3$ compound the restoration of the compressibility modulus parallel to the chain axis ($C_{33}$) does not exceed 50\% \cite{Poirier1995}.

As shown in the inset of  Fig.~\ref{fig.4}, the transverse mode data in zero magnetic field can be reasonably fitted by the expression (\ref{Cii}) using $|\Delta_q | \sim (1-T/T_{SP})^\beta$ with a critical exponent $\beta$ near 0.36 slightly  smaller than for CuGeO$_3$ (0.42) \cite{Poirier1995} and (TMTTF)$_2$PF$_6$ (0.49) \cite{Poirier2012}; a similar fit for the longitudinal mode in Fig.~\ref{fig.3} is however not reliable because of a too much scatter in the data.

We close the section by considering the magnetic field dependence of the SP critical temperature, $T_{\rm SP}$(H), and check if it follows the mean-field prediction for the spin-Peierls transition \cite{Cross1979}. First, we must decide how the transition temperature is determined to get a unique $T_{\rm SP}$ from the microwave and elastic data. For the microwave experiment, we mentioned that the second temperature derivative of $\Delta\epsilon$ was the most appropriate criterion for the determination of the critical temperature, which  yielded $T_{\rm SP} \simeq$ 17.9 K. In the elastic experiment, $T_{\rm SP}$ is most easily determined from the first temperature derivative of the transverse data because of the sharpness of the softening anomaly (indicated by an arrow in Fig.~4); this gives a temperature just above the minimum value of the softening. The same criterion has thus been applied to the longitudinal data (Fig.~3). In zero field, these procedures give an average temperature, $T_{\rm SP}$(0) = 17.9 $\pm$ 0.1 K, a value falling within  the range 17.1 - 18 K found from  other techniques \cite{Lumsden1999,Huizinga1979,Weckhuysen2001}. In the framework of mean-field theory of quasi-one-dimensional SP systems \cite{Cross1979,Azzouz1996}, the SP temperature is predicted to decay quadratically with the magnetic field according to the expression,
\begin{equation}
\frac{T_{\rm SP}(H)}{T_{\rm SP}(0)} = \left[1 - 14.4{\left(
\frac{g\mu_BH}{ 4\pi T_{\rm SP}(0)}\right)}^2\right],
\label{Th}
\end{equation}

\begin{figure}[H,h]
\includegraphics[width=8.5cm]{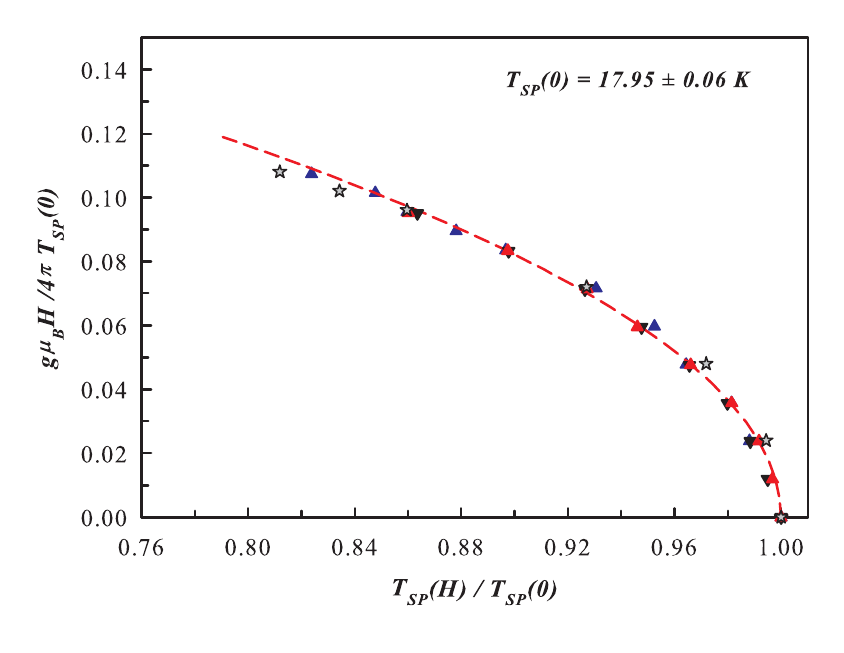} \caption{(Color online)
Magnetic field dependence of the reduced SP transition
temperature $T_{\rm SP}(H)/T_{\rm SP}(0)$ for MEM(TCNQ)$_2$. Experiments: microwave (stars); ultrasound (triangles), longitudinal(blue and black), transverse (red). The dashed line is the mean-field prediction from Ref.\cite{Cross1979}.}\label{fig.5}
\end{figure}
We present in Figure 5 the part of the magnetic phase diagram representing the D$-$U phase boundary given by Eq.~\ref{Th}. The results of mean-field calculations is plotted with a $g$ value of 2.0 \cite{Weckhuysen2001}. Experiments have been performed up to 18 Teslas for microwave and  longitudinal ultrasonic probes and up to 16 Teslas for transverse ultrasound experiments. All the data points agree with the theoretical prediction up to the highest field values. Up to 18 Tesla, however, our data fail to reveal the  D $-$high-field magnetic phase boundary for field scans at fixed temperatures, neither did they indicate the proximity of a critical point, as previously reported in magnetization close  to 18 Tesla \cite{Weckhuysen2001}. As far as the D$-$U boundary is concerned, MEM(TCNQ)$_2$ shows field profile similar to the one found for the organic Fabre salt (TMTTF)$_2$PF$_6$ \cite{Poirier2012}.

\section{CONCLUSION}

We have investigated charge, spin and lattice effects in the spin-Peierls state of the organic material MEM(TCNQ)$_2$ and discussed them in relation with other SP compounds, especially the Fabre salt (TMTTF)$_2$PF$_6$. Charge effects were studied by the microwave probe, which show an enhancement for the dielectric response at the onset of the SP state. Such measurements turn out to be relatively rare on these kind of systems, and few theoretical tools exist to account for the coupling of the charge degrees of freedom to the elastic strain (lattice) in  SP systems supposed to involve almost exclusively  spin degrees of freedom. Differences noted from (TMTTF)$_2$PF$_6$, in particular concerning the critical behaviour of the dielectric constant could be attributed to the 3D nature of the lattice fluctuations and the absence of a CO transition preceding the SP transition in MEM(TCNQ)$_2$. Magneto-elastic effects were revealed by characteristic anomalies appearing on the ultrasonic velocity of two elastic modes. These anomalies can be described with the aid of a Landau free energy model that includes a linear-quadratic coupling term between the elastic strain and the magnetic gap;  a situation that  again contrasts with (TMTTF)$_2$PF$_6$ for which biquadratic coupling is rather considered. Globally, all these effects aim at a conventional SP system like the other inorganic and organic compounds, as far as the critical behavior and the magnetic phase diagram are concerned. The order parameter critical exponent can be associated to a one component  universality class  in three dimensions, as to the field dependence of the D-U boundary of the phase diagram, it   is found to agree with the mean-field prediction for a quasi-one-dimensional system.

\acknowledgments{The authors acknowledge the technical support of Mario Castonguay and thank  Pascale Foury-Leylekian for discussion and comments.  This work was supported by grants from the Fonds Qu\'eb\'ecois de la Recherche sur la Nature et les Technologies (FQRNT) and from the Natural Science and Engineering Research Council of Canada (NSERC).}


\begin{thebibliography}{23}

\bibitem{Bray1973} J.W. Bray, H.R. Hart Jr., L.V. Interrante, L.S. Jacobs, J.S. Kasper, G.D. Watkins and S.H. Wee, Phys. Rev. Lett. \textbf{35}, 744 (1973).
\bibitem{Bray1983} J.W. Bray, L.V. Interrante, I.S. Jacobs and J.C. Bonner, \textit{Extended linear Chain Compounds} (Plenum, New York, 1983) Vol.~3, pp. 353-415.
\bibitem{Hase1993} M. Hase, I. Terasaki, K. Uchinokura, M. Tokunaga, N. Miura and H. Obara, Phys. Rev. B \textbf{48}, 9616 (1993).
\bibitem{Lumsden1998} M.D. Lumsden, B.D. Gaulin and H. Dabkowska, Phys. Rev. B \textbf{57}, 14097 (1998).
\bibitem{Lumsden1999} M.D. Lumsden and B.D. Gaulin, Phys. Rev. B \textbf{59}, 9372 (1999).
\bibitem{Cross1979} M.C. Cross, Phys. Rev. B \textbf{20}, 4606 (1979).
\bibitem{Liu1993} Q. Liu, S. Ravy, J.P. Pouget and C. Bourbonnais, Synthetic Metals \textbf{56}, 1840 (1993).
\bibitem{Dumoulin1996} B. Dumoulin, C. Bourbonnais, S. Ravy, J.P. Pouget and C. Coulon, Phys. Rev. Lett. \textbf{76}, 1360 (1996).
\bibitem{Pouget82} J. P. Pouget,R. Moret, R. Comes, K. Bechgaard, J.-M. Fabre and L. Giral, Mol. Cryst. Liq. Cryst. {\bf 79}, 129 (1982).
\bibitem{Creuzet1987} F. Creuzet, C. Bourbonnais, L.G. Caron, D. Jerome and K. Bechgaard, Synth. Met. \textbf{19}, 289 (1987).
\bibitem{Foury04} P. Foury-Leylekian, D. Le Bolloc'h, B. Hennion, S. Ravy, A. Moradpour, and J.-P. Pouget, Phys. Rev. B {\bf 70}, 180405 (2004).
\bibitem{Chow98} D. S. Chow,  P. Wzietek,  D. Fogliatti, B. Alavi,  D. J. Tantillo,  C. A. Merlic,  and S. E. Brown, Phys. Rev. Lett. {\bf 81}, 3984 (1998).
\bibitem{Pouget2006} J.-P. Pouget, P. Foury-Leylekian, D. Le Booloc'h, B. Hennion, S. Ravy, C. Coulon, V. Cardoso and A. Moradpour, J. Low Temp. Phys \textbf{142}, 147 (2006).
\bibitem{Coulon82} C. Coulon, P. Delhaes,  S. Flandrois, R. Lagnier, E. Bonjour, and J.M. Fabre, J. Physique (Paris), {\bf 43}, 1059 (1982).
\bibitem{Chow00} D. S. Chow, F. Zamborszky, B. Alavi, D.J. Tantillo, A. Baur,  C. A. Merlic and S. E. Brown, Phys.  Rev.  Lett. {\bf 85}, 13609 (2000).
\bibitem{Nad2000} F. Nad, P. Monceau, C. Carcel and J.M. Fabre, Phys. Rev. B \textbf{62}, 1753 (2000).
\bibitem{Coulon07} C. Coulon, G. Lalet, J.-P. Pouget, P. Foury-Leylekian, A. Moradpour, and J.-M. Fabre, Phys. Rev. B {\bf 76}, 085126 (2007).
\bibitem{Langlois2010} A. Langlois, M. Poirier, C. Bourbonnais, P. Foury-Leylekian, A. Moradpour and J.-P. Pouget, Phys. Rev.B \textbf{81}, 125101 (2010).
\bibitem{Poirier2012} M. Poirier, A. Langlois, C. Bourbonnais, P. Foury-Leylekian, A. Moradpour and J.-P. Pouget, Phys. Rev. B \textbf{86}, 085111 (2012).
\bibitem{Azzouz1996} M. Azzouz and C. Bourbonnais, Phys. Rev. B \textbf{53}, 5090 (1996).
\bibitem{Poirier1995} M. Poirier, M. Castonguay, A. Revcolevschi and G. Dhalenne, Phys. Rev. B\textbf{52}, 16058 (1995).
\bibitem{Bodegom1981} B. van Bodegom, Acta Crystallogr., Sect B: Struct. Crystallogr. Cryst. Chem. \textbf{37}, 863 (1981).
\bibitem{Huizinga1979} S. Huizinga, J. Kommandeur, A.A. Sawtzky, T.T. Thole, K. Kopinga, W.J.M. de Jonge and J. Roos, Phys. Rev. B \textbf{19}, 4723 (1979).
\bibitem{Bodegom1981a} B. van Bodegom, B.C. Larson and H.A. Mook, Phys. Rev. B \textbf{24}, 1520 (1981).
\bibitem{Blundell1997} S.J. Blundell, F.L. Pratt, P.A. Pattenden,M. Kurmoo, K.H. Chow, S. Tagagi, Zh. Jest$\ddot{a}$dt and W. Hayes, J. Phys.: Condens. Matter \textbf{9}, L119 (1997).
\bibitem{Weckhuysen2001} L. Weckhuysen, S.V. Demishev, A.V. Semeno, J. Vanacken, L. Trappeniers, A.A. Pronin and V.V. Moshchalkov, Europhys. Lett. \textbf{53}, 667 (2001).
\bibitem{Buravov} L. Buravov and J. F. Shchegolev, Prib. Tekh. Eksp. \textbf{2}, 171 (1971).
\bibitem{Morrow1980} M. Morrow, W.N. Hardy, J.F. Carolan, A.J. Berlinsky, L. Weller, V.K. Gujral, A. Janossy, K. Holczer, G. Mihaly, G. Gr\"{u}ner, S. Huizinga, A. Verwey, and G.A. Sawatsky, Can. Jour. Phys. \textbf{58}, 334 (1980).
\bibitem{Visser1983} R.J. Visser, S. Oostra, C. Vettier and J. Voiron, Phys. Rev. B \textbf{28}, 2074 (1983).
\bibitem{Pouget1989} J. Pouget, in \textit{Low-Dimensional Electronic Properties of Molybdenum Bronzes and Oxides}, edited by C. Schlenker (Kluwer Academic, Netherlands, 1989), pp. 87-157.

\end{thebibliography}
\end{document}